\documentclass{article}
\usepackage[utf8]{inputenc}
\usepackage{longtable}
\usepackage{booktabs}
\usepackage{setspace}
\usepackage{graphicx}
\usepackage[table]{xcolor}

\usepackage{amsmath, amsthm, amssymb}
\usepackage{natbib}

\usepackage{mdframed}
\usepackage{enumitem}

\usepackage[margin=1.5in]{geometry}
\usepackage{hyperref}
\hypersetup{colorlinks,citecolor=blue,urlcolor=blue,linkcolor=blue}
\usepackage{pgfplots}

\newcommand{\authorblock}[1]{\begin{tabular}{@{}c@{}}#1\end{tabular}}

\usetikzlibrary{arrows.meta}
\usetikzlibrary{backgrounds}
\usepgfplotslibrary{patchplots}
\usepgfplotslibrary{fillbetween}

\newcommand{\trialverify}{\emph{TrialProbe}}

\newcommand{\p}[1]{\left(#1\right)}

\newcommand{\cb}[1]{\left\{#1\right\}}
\newcommand{\simiid}{\,{\buildrel \text{iid} \over \sim\,}}
\newcommand{\EE}[2][]{\mathbb{E}_{#1}\left[#2\right]}
\DeclareMathOperator*{\argmax}{argmax}

\author{ 
  \begin{tabular}{c@{\qquad}c} 
    \authorblock{Ethan Steinberg \\ \texttt{ethanid@stanford.edu}} & 
    \authorblock{Nikolaos Ignatiadis \\ \texttt{ignat@stanford.edu}} \\ 
    \\ 
    \authorblock{Steve Yadlowsky \\ \texttt{yadlowsky@google.com}} & 
     \authorblock{Yizhe Xu \\ \texttt{yizhex@stanford.edu}} \\
     \\ 
     \multicolumn{2}{c}{\authorblock{Nigam H. Shah \\ \texttt{nigam@stanford.edu }}}\\ &
  \end{tabular}  
}
\title{Using public clinical trial reports to probe observational study methods}
\date{August 2022}

\begin{document}

\maketitle

\begin{abstract}
Observational studies are valuable for estimating the effects of various medical interventions, but are notoriously difficult to evaluate because the methods used in observational studies require untestable assumptions. This lack of intrinsic verifiability makes it difficult both to compare different observational study methods and to trust the results of any particular observational study. Reference sets, which are a source of relationships that are believed to be true, are often used to evaluate observational study methods. In this work, we propose TrialProbe, an approach for evaluating methods used in observational studies using ground truth sourced from clinical trial reports. We process published trial reports into a denoised collection of known causal relationships that, given a dataset, can then be used to estimate the precision and recall of various observational study methods. We then use TrailProbe to evaluate three common observational study methods in terms of their ability to identify the known causal relationships from a large national insurance claims dataset. We find that adjusting for confounding significantly improves our ability to correctly recover effects, with propensity score matching performing particularly well at detecting large effects. TrialProbe is made freely available for others to evaluate future observational study methods.
\end{abstract}

\section{Introduction}

Observational studies are valuable for estimating causal relationships in medical settings where randomized trials are not feasible due to either ethical or logistical concerns \citep{pmid25005647}. The primary issue with observational studies is that observational data intrinsically contains confounding variables that are related to both treatment assignments and outcomes and induce artificial relationships between them that prevent the identification of true causal relationships. Causal inference methods for observational studies, such as adjusting for confounders, are widely applied in comparative effectiveness research in order to reveal unbiased estimates of treatment effects. However, these methods are limited by their reliance on untestable assumptions, such as the absence of unmeasured confounding and non-informative censoring, which cannot be verified based on observational data alone \citep{Berger2017-qy}. These untestable assumptions make it difficult to evaluate the performance of observational study methods, which is essential for verifying reliability of these techniques as well as determining the relative accuracy of different methods. This hampers research, by making it more difficult to develop more effective methods, and hinders practice, as clinicians are hesitant to use evidence generated from observational data even in situations where clinical trial derived evidence is not available \citep{Hampson2018-ld, Klonoff2020-hu}.

One common approach for evaluating observational study methods is through reference sets \citep{Schuemie2013-nh, Schuemie2020-aw, Ryan2013-kb}. A reference set is a collection of relationships about the effects of treatments that are independently verified, and treated as ground truth against which the ability of an observational study method to identify those effects from available data can be quantified. Even in an optimal situation, it is not expected that any observational method will reproduce the ground truth in the reference set because the observational data usually comes from a different population than the population used to collect the ground truth \citep{Thompson2021-op}. Identification of a known relationship might fail for example because the academic medical center population used in an RCT might differ drastically from the general population available in observational data sets. However, a common assumption is that while the exact effect might differ, the effect identified in the observational data and the original ”ground truth” should be correlated and good observational methods should on average have greater correspondence with the provided ground truth \citep{Camerer2018-cd}. There have been several proposed approaches to create reference sets , the most prominent of which rely on either FDA labels or expert knowledge to declare known relationships between drugs and outcomes \citep{Ryan2013-kb}. However, the quality of the resulting reference sets has been questioned, as some of the evidence underpinning these reference sets is relatively low quality \citep{Coloma2013-nu, Schuemie2013-nh, Schuemie2020-aw}.

In this work, we introduce TrialProbe, a new principled approach for constructing a reference set from publicly available clinical trial reports at www.clinicaltrials.gov. We make three main improvements on previous approaches to create reference sets. First, we explicitly focus the side effects of active comparator study designs where one drug is directly compared to another drug as those are much easier to connect to potential observational study designs \citep{Hernan2016-ja}. Second, we use an empirical Bayes analysis to compute denoised ground truth effect estimates. Third, we explicitly filter for effects with large effect sizes to reduce the influence of noise caused by differences between RCTs and observational studies.

We then use the resulting reference set to evaluate several common observational study methods in terms of their ability to identify the true relationships from a large national insurance claims dataset – the Optum Clinformatics Data Mart. We also introduce metrics for quantifying the quality of such identification when comparing the results from different observational study methods and provide reliability estimates at different detected effect sizes. We find that we can reproduce a significant fraction of the reported effects and find that propensity score based adjustment significantly improves our ability to reproduce effects. We also compare our reference set to two existing reference sets to show how TrialProbe enables better comparison of methods.

\section{\trialverify}

In this Section we describe the \trialverify~approach. Concretely, we describe the data source of the clinical trial reports (ClinicalTrials.gov), the processing of the raw data to a curated dataset of $M=19,822$ unique drug/drug adverse event comparisons, as well as the statistical approach that we propose for comparing observational study methods.

\subsection{The primary data source: ClinicalTrials.gov}

ClinicalTrials.gov serves as a public repository for clinical trials
carried out in the United States and abroad. The database contains
pre-registration information, trial status, and results as provided by
researchers conducting the trials. Many clinical trials are
legally required to report results to ClinicalTrials.gov within 1 year of study completion~\citep{devito2020compliance}. In this work we use the June 4, 2020
version of the database which includes 33,701 clinical trials.

\subsection{Extracting trials with an active comparator
design}
\label{sec:extracting}

We focus on drug versus drug active comparator clinical trials, which
evaluate one drug directly against another. The reason is that such comparisons are easier to conduct in the context of a observational study design. In contrast, placebo or standard of
care based trials are more difficult to work with because there is no
clear corresponding case-control observational study that can be used to
estimate effects.

The results section of each active comparator clinical trial record
consists of a set of intervention arms as well as the primary outcomes
and adverse events associated with each arm. The primary outcomes and side
effects are all specified in natural language and must be mapped to
standardized terminologies. We discard the
primary outcomes because it is difficult to consistently map them to
electronic healthcare data sources due to a wide diversity of measurements and a lack of
standardized terminology. We instead focus on the adverse event because
it is specified using MedDRA terminology and mappings to corresponding
condition codes are available for healthcare data sources. We obtain a
standardized version of these adverse outcomes by mapping them to ICD10
using the dictionary mappings contained within UMLS 2019AB.

The drug mentions in the ClinicalTrials.gov records are specified in an
ad-hoc manner in terms of brand names, ingredients, dosages and/or more
specialized names. As a preliminary step, we filter out all treatment
arms with fewer than 100 patients as trials of that size frequently do
not have enough power to obtain statistical significance. We then use
the RxNorm API to transform the text descriptions of drugs into RxNorm
ingredient sets. We require at least 50\% of the tokens to match in
order to avoid false positives. Treatment arms with more than one
ingredient (due to either containing multiple drugs or drugs with
multiple active ingredients) are also filtered out. As an additional
quality control step, we remove intervention arms that contain plus (``$+$'')
signs in their names that usually indicate combination treatments that
RxNorm is not always able to detect and map to ingredients correctly. Finally, we map those RxNorm ingredient sets to ATC codes (cite ATC) so that we can find the corresponding drugs more easily in our ATC annotated observational data.

One important feature of ClinicalTrials.gov data is that it often
contains records where the same drug-drug comparisons have been tested
in multiple trials. We aggregate side effect event counts and
participant counts for trials with identical drug combinations and
outcome measurements. Similarly, we also aggregate counts across arms
where the same drug was evaluated with different dosages. This
aggregation procedure has the dual purposes of strengthening the
reliability of consistent true effects while helping to down-weigh trials
with conflicting effects.

We also note that in an active comparator design there is typically no concrete choice for the baseline arm (in contrast to e.g., placebo or standard of care trials)---the role of the two arms is \emph{symmetric}. To express this symmetry, we reorder all pairs of drugs under comparison (for each adverse event) in such a way that the sample odds ratio is $\leq 1$.

At the end of this process, we have compiled $M=19,822$
unique drug versus drug treatment adverse event comparisons. The summarized data for the $i$-th entry comprises of the ICD10 code of the adverse event, the ATC code of the two drugs being compared, as well as the contingency table $Z_i$:
\begin{equation}
\label{eq:contingency}
Z_i = \begin{array}{|lcc|}
\hline
& \text{Drug A} & \text{Drug B} \\
 \text{Number of patients with the adverse event} & X_{A,i} & X_{B,i} \\
 \text{Number of patients without the adverse event} & Y_{A,i} & Y_{B,i} \\
\hline
\end{array}
\end{equation}
Below we describe our concrete statistical proposal for leveraging the above dataset to compare observational study methods.

\subsection{Empirical Bayes ranking}
\label{subsec:empirical_bayes}

In this section we develop an approach to ranking all the drug versus drug treatment adverse event comparisons in the \trialverify~dataset of Section~\ref{sec:discordant} that accommodates the following desiderata: First, comparisons for which we have \emph{strong} evidence of a \emph{large} odds ratio are prioritized. Second, comparisons with  odds ratios $\approx 1$, or with insufficient numerical evidence for a large odds ratio (e.g.,  few patients in the contigency table $Z_i$~\eqref{eq:contingency} or few patients with adverse events for both drugs) are down-ranked. The constructed ranking will play a key role in the evaluation procedure described below in Section~\ref{sec:discordant}.

Our ranking approach follows a tradition of methodological developments based on hierarchical modeling combined with an empirical Bayes analysis~\citep{aitkin1986statistical, laird1989empirical, henderson2016making, gu2022invidious}. We model the likelihood for the log-odds ratio $\omega_i$ of the $i$-th comparison (with contingency table~\eqref{eq:contingency}) through the non-central hypergeometric distribution, that is,
\begin{equation}
\label{eq:conditional_likelihood}
L_i(\omega_i)=\frac{\binom{X_{A,i} + Y_{A,i}}{X_{A,i}}
\binom{X_{B,i} + Y_{B,i}}{X_{B,i}} \mathrm{e}^{\omega_i X_{A,i}}}{\sum_{t}\binom{X_{A,i} + X_{B,i}}{t}\binom{X_{B,i} + Y_{B,i}}{X_{A,i} + X_{B,i} - t} \mathrm{e}^{\omega_i t}}.
\end{equation}
The likelihood $L_i(\omega_i)$ for the analysis of $2 \times 2$ contingency tables has been proposed by \citet*{van1993bivariate, efron1996empirical, sidik2008estimation}, and~\citet*{stijnen2010random}, and is derived by conditioning on the margins of the table $Z_i$---in entirely the same way as in the derivation of Fisher's exact test. 

In our hierarchical approach, we further model the $\omega_i$ as exchangeable random effects, independent of the margins of $Z_i$, with:
\begin{equation}
\label{eq:eb_model}
\omega_i\, (i=1,\dotsc,M) \, \simiid\,  G, \;\;\; \text{G} \text{ is a symmetric distribution around the origin}.
\end{equation}
In contrast to a Bayesian approach, we do not posit knowledge of $G$, but instead follow the empirical Bayes paradigm and estimate $G$ based on the data $Z_1,\dotsc,Z_M$ as follows:
\begin{equation}
\label{eq:NPMLE}
    \widehat{G} \in \argmax_G \cb{ \sum_{i=1}^M \log\p{ \int L_i(\omega) dG(\omega)}\;:\; \text{G} \text{ is a symmetric distribution around }0}.
\end{equation}
\eqref{eq:NPMLE} is an optimization problem over all symmetric distributions $G$ and the objective is the marginal log-likelihood---each component likelihood $L_i(\cdot)$~\eqref{eq:conditional_likelihood} is integrated with respect to the unknown $G$. The estimator $\widehat{G}$ is the nonparametric maximum likelihood estimator (NPMLE) of~\citet{kiefer1956consistency}, and has been used for contigency tables, e.g., by~\citet{van1993bivariate}. We note that  in contrast to previous works (e.g.~\citet{van1993bivariate}), we also enforce  symmetry of $G$ around $0$ in~\eqref{eq:eb_model}, \eqref{eq:NPMLE}. The reason is that, as explained in Section~\ref{sec:extracting}, our active comparator design setting is symmetric with respect to the drugs under comparison.

\begin{figure}
    \centering
    \begin{tabular}{ll}
    a) & b) \\
    \resizebox{0.5\linewidth}{!}{\input{figures/trialverify_distributions.tex}} & \includegraphics[width=0.5\linewidth]{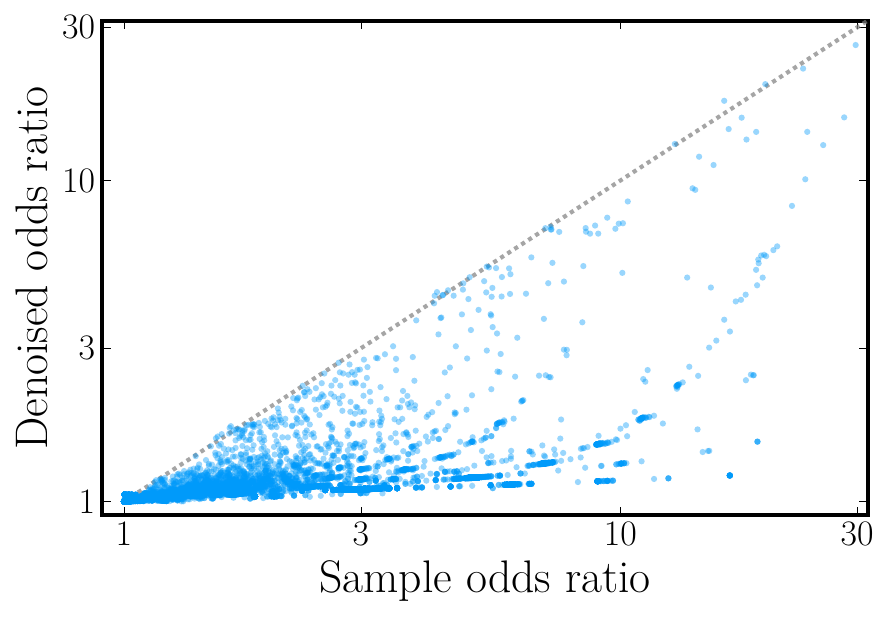} \\
    \end{tabular}
    \caption{a) Distribution function of drug versus drug adverse event odds ratios in \trialverify. $\widehat{G}$ is estimated via nonparametric maximum likelihood as in~\eqref{eq:NPMLE}, while the dashed curve is the empirical distribution of sample odds ratios. b) Denoised vs. raw odds ratios. Denoising~\eqref{eq:EB_shrinkage} is done by computing the posterior mean of the log odds ratio given the data for the $i$-th comparison and the estimated $\widehat{G}$.}
    \label{fig:eb}
\end{figure}

Figure~\ref{fig:eb}a) shows the estimated distribution function $\widehat{G}$~\eqref{eq:NPMLE} based on the TrialProbe dataset (in terms of odds ratios $\exp(\omega_i)$, but with a logarithmic $x$-axis scale), as well as the empirical distribution of sample odds ratios.\footnote{Computed with a pseudocount adjustment to deal with zero cell counts, that is, $\exp(\widehat{\omega}^{\text{sample}}_i)= \p{(X_{A,i}+0.5)/(Y_{A,i}+1)}\big/\p{(X_{B,i}+0.5)/(Y_{B,i}+1)}.$} We observe that even though the sample odds ratios are quite spread out, the NPMLE $\widehat{G}$ is substantially more concentrated around odds ratios near $0$. This is consistent with the intuition that for an active comparator design study, side effects will often be similar for the two drugs under comparison (but not always). 
Finally, to rank the drug versus drug treatment adverse event comparisons we use the intuitive approach suggested e.g., by~\citet{aitkin1986statistical}: We use the estimated \smash{$\widehat{G}$} to compute denoised point estimates of the odds ratios via
\begin{equation}
\label{eq:EB_shrinkage}
 \widehat{\omega}_i^{\text{EB}}= \EE[\widehat{G}]{ \omega_i \mid Z_i} = \frac{\int \omega L_i(\omega) d\widehat{G}(\omega)}{\int  L_i(\omega) d\widehat{G}(\omega)},
\end{equation}
and we also rank the comparisons by  $\widehat{\omega}_i^{\text{EB}}$. Figure~\ref{fig:eb}b) plots $\exp(\widehat{\omega}_i^{\text{EB}})$ against the sample odds ratios. We observe that the rule  $\widehat{\omega}_i^{\text{EB}}$ automatically shrinks most sample odds ratios toward $1$, while rigorously accounting for varying effective sample of each comparison (so that shrinkage toward $1$ is heterogeneous). Table \ref{reference_set_entries} gives 10 example entries from our refernece Set.

\begin{table}[]
    \centering
    \begin{tabular}{ p{3cm}  p{3cm}  p{3cm} p{2cm} p{2cm} }
 Adverse Event (ICD10) & Drug A (ATC) & Drug B (ATC) & Contingency Table & Denoised odds ratio \\
\hline

Acne \newline (L70) & Cisplatin \newline (L01XA01) & Panitumumab \newline (L01XC08) &
        $\lceil 1 \hspace{0.2cm} 155 \rceil$ \newline
        $\lfloor 147 \hspace{0.2cm} 12 \rfloor$
     & 1754.58\\
Alopecia \newline (L65.9) & Atezolizumab \newline (L01XC32) & Docetaxel \newline (L01CD02) &
        $\lceil 8 \hspace{0.2cm} 743 \rceil$ \newline
        $\lfloor 257 \hspace{0.2cm} 456 \rfloor$
     & 41.42\\
Alopecia \newline (L65.9) & Avelumab \newline (L01XC31) & Docetaxel \newline (L01CD02) &
        $\lceil 3 \hspace{0.2cm} 390 \rceil$ \newline
        $\lfloor 97 \hspace{0.2cm} 268 \rfloor$
     & 32.15\\
Neutropenia \newline (D70) & Avelumab \newline (L01XC31) & Docetaxel \newline (L01CD02) &
        $\lceil 2 \hspace{0.2cm} 391 \rceil$ \newline
        $\lfloor 58 \hspace{0.2cm} 307 \rfloor$
     & 23.47\\
Nausea \newline (R11.0) & Insulin Glulisine \newline (A10AB06) & Lixisenatide \newline (A10BJ03) &
        $\lceil 8 \hspace{0.2cm} 587 \rceil$ \newline
        $\lfloor 75 \hspace{0.2cm} 223 \rfloor$
     & 20.74\\
Alopecia \newline (L65.9) & Pemetrexed \newline (L01BA04) & Docetaxel \newline (L01CD02) &
        $\lceil 5 \hspace{0.2cm} 390 \rceil$ \newline
        $\lfloor 54 \hspace{0.2cm} 197 \rfloor$
     & 18.01\\
Rash \newline (R21) & Interferon Beta-1a \newline (L03AB07) & Alemtuzumab \newline (L04AA34) &
        $\lceil 9 \hspace{0.2cm} 178 \rceil$ \newline
        $\lfloor 174 \hspace{0.2cm} 202 \rfloor$
     & 16.61\\
Nail Disorder \newline (L60) & Atezolizumab \newline (L01XC32) & Docetaxel \newline (L01CD02) &
        $\lceil 2 \hspace{0.2cm} 749 \rceil$ \newline
        $\lfloor 39 \hspace{0.2cm} 674 \rfloor$
     & 12.87\\
Tinnitus \newline (H93.1) & Panitumumab \newline (L01XC08) & Cisplatin \newline (L01XA01) &
        $\lceil 18 \hspace{0.2cm} 141 \rceil$ \newline
        $\lfloor 98 \hspace{0.2cm} 58 \rfloor$
     & 12.19\\
Dysgeusia \newline (R43.2) & Brimonidine \newline (D11AX21, S01EA05) & Brinzolamide \newline (S01EC04) &
        $\lceil 2 \hspace{0.2cm} 453 \rceil$ \newline
        $\lfloor 38 \hspace{0.2cm} 422 \rfloor$
     & 11.83\\

\end{tabular}

    \caption{Example TrialProbe entries}
\label{reference_set_entries}
\end{table}

\subsection{Evaluation: Discordant signs}
\label{sec:discordant}

As explained previously, there are many possible reasons why the results of a observational assessment of a causal effect may not match the results of a clinical trial. We propose to handle this by only looking at the estimated effect direction for effects which are known to be large. The basic premise of our approach is the following.

\begin{mdframed}
Suppose the following hold for the comparison of two drugs with respect to an adverse event:
\begin{enumerate}
    \item In the clinical trial report, there is \emph{strong} evidence that $\omega_A \gg \omega_B$, that is, there is strong evidence that the adverse event rate under drug A is \emph{ substantially larger} compared to drug B.
    \item The observational study method yields a significant p-value for the null hypothesis that both drugs have the same adverse event rate.
    \item According to the observational study method, drug B leads to a higher adverse event rate compared to drug A, that is, the direction of the effect is the opposite compared to the clinical trial evidence.
\end{enumerate}
Then, we may suspect that the observational study method yielded misleading evidence.
\end{mdframed}
We instantiate the above framework as follows. Let $\mathcal{S} \subset \trialverify$ be a set of drug vs. drug adverse event comparisons. Furthermore, given an observational study method $\mathcal{O}$, let $\mathcal{R}(\mathcal{O}) \subset \trialverify$ be the set of comparisons such that the observational study returns a p-value $\leq 0.05$. We then define the Discordant Sign Rate, as:
\begin{equation}
    \text{DSR}(\mathcal{S}, \mathcal{O}) = \frac{ \#\cb{i\in \mathcal{S}\cap \mathcal{R}(\mathcal{O}): \text{Direction of effect for $i$ disagrees between } \mathcal{S},\mathcal{O}}}{ \# \mathcal{S}\cap \mathcal{R}(\mathcal{O})}.
\end{equation}

The primary question is how to construct $\mathcal{S}$. The primary tradeoff with $\mathcal{S}$ is that a larger reference set will have more datapoints and thus be lower variance, but at the cost of potentially including weaker effects that might not be as reliable. 

As there is no optimal $\mathcal{S}$, we evaluate on every possible option $(\mathcal{S}_t)_{t \geq 1}$ with:
\begin{equation}
    \label{eq:parameterized_family}
    \mathcal{S}_t = \cb{i \in \trialverify\,:\, \exp(\widehat{\omega}_i^{\text{EB}}) \geq t}.
\end{equation}

For every $\mathcal{S}_t$, we compute two metrics of interest: the fraction of statistically significant results that have the correct sign and the fraction of reference set entries correctly recovered (as in being marked statistically significant with the correct sign). The fraction with the correct sign gives an indication of how reliable an observational method is and the fraction recovered gives an indication of often an observational method misses significant effects.

\section{Case study on Optum Clinformatics}

We illustrate the value of \trialverify by investigating the performance of three
observational study methods on the Optum Clinformatics Data Mart 8.0
medical claims dataset~\citep{clininformatics}. The Optum dataset is a large US commercial claims dataset containing over 88 million patients and is frequently
used for observational studies. Our reference set contains 19,822 entries that specify to what degree
one drug is causally related to a side effect relative to another drug.

For each entry, we perform a new-user case-control observational study that
compares the side effects in control patients, who were started on the
control drug, to the side effects in intervention patients, who were started
on the intervention drug. All cohorts are constructed systematically
using the first drug reimbursement claim for either of the two drugs as
the index time. Patients with a prior event or an event at the index
time are excluded. At most 100,000 patients are sampled for each drug
with a minimum required patient count of 100. Side effects are measured
until each record is censored (as indicated by the end of their
healthcare enrollment in the Optum dataset). Figure 3 gives an overview
of our experimental setup.

\subsection{Methods}
For each of the cohorts constructed above using the Optum data, we
evaluate three methods for estimating treatment effects on the hazard
ratio scale, one providing an unadjusted estimate, and two that adjust for the presence of confounding. Our unadjusted method is a
univariate Cox model, using the exponentiated coefficient of the
treatment as the effect estimate. Our two adjusted methods are a propensity
score matched Cox model and an inverse propensity score weighted Cox model. The propensity score is estimated using
logistic regression on a low-dimensional representation of the patient's
history obtained via a procedure by~\citet{steinberg2021language}.
When performing matching, we use a 1:1 greedy matching algorithm on the
logit scale with a caliper of 0.1. Once a matched cohort is chosen, the
hazard ratio is estimated using a Cox regression by modeling the
survival outcome as a function of the treatment status in the cohort.

\subsection{Results}

We then compute the fraction of significant results that have the correct sign and the fraction of reference set entries covered for each method and each subset $S_t$ of TrialProbe that only contains effects that have an odds ratio threshold greater than $t$. Figure \ref{fig:eval} provides the performance of each of our three methods on these two metrics of interest.

\begin{figure}
    \centering
    \begin{tabular}{ll}
    a) & b) \\
    \resizebox{0.5\linewidth}{!}{\input{figures/trialverify_sign_rate.tex}} &
    \resizebox{0.5\linewidth}{!}{\input{figures/trialverify_found.tex}} 
    \end{tabular}
    \caption{a) Fraction of significant results with the correct sign as a function of the odds ratio threshold.
 b) Fraction of correctly recovered entries as a function of the odds ratio threshold.}
    \label{fig:eval}
\end{figure}

We find that the fraction correct and fraction recovered drop as we decrease the odds ratio threshold. This reflects how smaller effects are harder to find correctly in observational data. As expected, the fraction correct does seem to drop to noise (50\% correct) as the odds ratio threshold approaches 1.0 and effects become very small.

We also find that the adjusted Cox models have a significantly higher correctness rate compared to the unadjusted models at the cost of recovering fewer reference set effects in total. The comparison between propensity score matching and inverse propensity score weighting is more complicated. Propensity score matching recovers more effects, but suffers from reduced accuracy for weaker effects.

  \subsection{Comparison to OMOP and EU-ADR reference sets}

As a comparison, we perform the same evaluation using the OMOP and
EU-ADR reference sets. The OMOP and EU-ADR reference sets significantly
differ from \trialverify~in that they have negative
controls with a presumed zero effect and positive controls with a
presumed non-zero effect. This is comparable to our metrics when using an odds ratio threshold of 1.

\begin{table}[]
    \centering
    
    \begin{tabular}{|l|c|c|c|c|}
\hline
\multicolumn{1}{|c|}{} & \multicolumn{2}{c|}{OMOP} & \multicolumn{2}{c|}{EU-ADR}\\
\cline{2-5}
\multicolumn{1}{|c|}{Method} & \% Correct & \% Recovered & \% Correct & \% Recovered \\
\hline
Unadjusted Cox   &   0.513 & 0.445 & 0.619 & 0.356 \\
Propensity Score Matched Cox &   0.527 & 0.185 & 0.538 & 0.095   \\
Inverse Propensity Score Weighted Cox & 0.590 & 0.4102 & 0.576 & 0.424 \\
\hline
\end{tabular}
    \caption{Performance of observational methods according to the EU-ADR and OMOP reference sets}
    \label{tab:omop_euadr}
\end{table}

Similar to prior work \citep{schuemie2013replication}, we had difficulties recovering that many effects, with low performance in both correctness and recovery rate in all of our methods.

\section{Discussion} 
We used clinical trial records from ClinicalTrials.gov to build a source of ground truth (i.e, a reference set) to evaluate observational study methods. We have shown how such a reference set can be constructed in a systematic manner that sets pre-specified rates of false positives. We also have demonstrated the value of our approach by showing how using this reference set can quantify the performance of eleven commonly used observational study methods.

Our approach has three advantages. First, it evaluates the effectiveness of methods in realistic scenarios on real observational data. Alternative approaches such as simulations or semi-simulated data require the creation of realistic simulated relationships . It is often difficult to determine whether or not those simulations provide a realistic confounding structure that is similar to observational data in practice \citep{schuler2017synth, 1707.02641}. Second, our approach provides high quality ground truth based on clinical trials that have meaningfully large effects and can be compared directly to corresponding observational studies. Prior reference sets rely on non-clinical trial sources that might be less reliable or have weaker relationships to potential observational studies. One indication of our reference set quality is that we are able to identify the relationships in it quite well using the Optum data, with relatively high precision and recall. Finally, our approach scales better than prior work, as we can take advantage of large sets of public clinical trial reports. This enables us to create hundreds of “known relationships” to quantify the performance estimates of the methods examined. This is a significant advantage compared to prior approaches that rely on evaluating inference methods using individual randomized trial datasets that can be difficult to acquire \citep{powers2018some}.

However, our approach has some limitations. First, we rely on an assumption that the average treatment effect seen in the clinical trials generalizes to the population available in the observational data. If there is a significant mismatch in population and there is a heterogeneous treatment effect, it is possible to see different effect directions in the observational data than the randomized trial even if the observational study methods are functioning correctly \citep{Rogers2021-en, Dahabreh2020-wv}. We somewhat account for this by only trying to reproduce the direction, which should be more stable across populations than the exact effect size. The use of odds ratios also somewhat adjusts for changing conditions as odds ratios are more stable in the presence of underreporting issues that are common in observational data sets. In addition, we would expect this type of error to lead to an underestimate of the performance of the observational study methods. A second limitation is that our approach is only able to evaluate methods for detecting average treatment effects because our ground truth is in the form of average treatment effects. We are simply unable to evaluate how effective methods can detect heterogeneous treatment effects. A third limitation is that the evaluation results will vary based on the underlying observational dataset used for identifying the known effects, the Optum Clinformatics Data Mart in our case. It is possible that some methods perform better in some observational datasets and worse in others. Care must be taken to only compare methods using the same observational dataset to ensure a fair comparison.

\section{Conclusion}
We propose an approach for evaluating observational study methods using clinical trial derived reference sets, and evaluate three commonly used observational study methods in terms of their ability to identify the known relationships using a commonly used claims dataset. We find that adjustment significantly improves the ability to correctly recover known relationships, with propesity score matching performing particularly well for detecting large effects.

We make TrialProbe, i.e. the reference set as well as the procedure to create it, freely available at https://github.com/som-shahlab/TrialProbe. TrialProbe is useful for developers of observational study methods for benchmarking their methods’ performance as well as for practitioners interested in knowing the expected performance of their methods of choice on the datasets available to them.

\section*{CRediT authorship contribution
statement}\label{credit-authorship-contribution-statement}
\begin{tabular}{ll}
Ethan Steinberg: & Conceptualization, Methodology, Software, Writing---original draft.\\
Nikolaos Ignatiadis: & Methodology, Software, Writing.\\
Steve Yadlowsky: & Methodology, Software, Writing.\\
Yizhe Xu: & Software, Writing.\\
Nigam H. Shah: & Writing---review \& editing,
Supervision, Funding acquisition.
\end{tabular}

\section*{Declaration of Competing
Interest}\label{declaration-of-competing-interest}

The authors declare that they have no known competing financial
interests or personal relationships that could have appeared to
influence the work reported in this paper.

\section*{Acknowledgments}\label{acknowledgments}

This work was funded under NLM R01-LM011369-05. GPU resources were
provided by Nero, a secure data science platform made possible by the
Stanford School of Medicine Research Office and Stanford Research
Computing Center. We would also like to thank Agata Foryciarz, Stephen
R. Pfohl, and Jason A. Fries for providing useful comments on the paper.

\bibliographystyle{plainnat}

\bibliography{references}

\end{document}